\documentstyle[11pt,epsfig]{article}
\begin{document}
\setlength{\baselineskip}{0.30in}
\newcommand{\beq}{\begin{equation}}
\newcommand{\eeq}{\end{equation}}
\newcommand{\bi}{\bibitem}

\begin{center}
\vglue .06in
{\Large \bf {Nonequilibrium Corrections to the Spectra of Massless
Neutrinos in the Early Universe -- Addendum.}
}
\bigskip
\\{\bf A.D. Dolgov
\footnote{Also: ITEP, Bol. Cheremushkinskaya 25, Moscow 113259, Russia.}
\footnote{e-mail: dolgov@tac.dk}, 
S.H. Hansen\footnote{e-mail: sthansen@tac.dk}
 \\[.05in]
{\it{Teoretisk Astrofysik Center\\
 Juliane Maries Vej 30, DK-2100, Copenhagen, Denmark
}}}
\\{\bf D.V. Semikoz\footnote{e-mail: semikoz@ms2.inr.ac.ru}}\\
{\it{Institute of Nuclear Research of the Russian Academy of Sciences\\
 60th October Anniversary Prospect 7a , Moscow 117312, Russia
}}\\[.40in]

\end{center}
\begin{abstract}

We repeat our previous calculation of the spectrum distortion of massless 
neutrinos in the early universe with a considerably better accuracy
and corrected for a missing numerical factor in one of the two ways 
of calculations presented in our paper~\cite{dhs}.
Now both ways of calculations are in perfect agreement and we essentially 
reproduce our old results presented in the abstract of the paper and 
used in the calculations of light element abundances.
We disagree with the criticism of our calculations presented in 
ref.~\cite{gnedin}.
\end{abstract}
\newpage

\section{Introduction}
In ref.~\cite{dhs} we accurately calculated the distortion of 
the spectrum of primordial massless neutrinos due to the interaction 
with hotter electrons and positrons in the primeval cosmic plasma.
Our results are in a reasonable agreement with the previous 
approximate or less accurate calculations (the list of references 
can be found in ref.~\cite{dhs}).
Recently a considerably different result was presented in ref.~\cite{gnedin}.
In view of that, we have redone our calculations with a better 
accuracy and also corrected for a missing numerical factor in 
one of the two different but equivalent ways of calculations 
presented in ref.~\cite{dhs} (see below). Due to this error
the results obtained by the two different methods were somewhat different.
Now both ways of calculations are in perfect agreement and we 
essentially reproduced our old results with a better 
precision\footnote{The results
quoted in the abstract of ref.~\cite{dhs} as well as the results used
in the calculation of $\Delta Y_{He}$ are the correct ones.}.

The increase in the energy density of neutrinos and the distortion 
of the spectrum of electronic neutrinos have little effect on light 
element abundance, but
it was recently pointed out that the corresponding increased fraction of
cosmic relativistic matter has an impact on the CMB anisotropies, 
which may be detectable by coming satellite experiments~\cite{gnedin, lopez}.

\section{Description of calculation procedure}

\subsection{Two ways of calculation}
In our previous paper on the subject~\cite{dhs} we  used two  
different  approaches for calculating the distortion
of the neutrino spectra.
We numerically solved the system of 
equations both for the full neutrino
distribution functions,  $f_{\nu _j}$, and for the deviations from equilibrium,
$\delta_j = (f_{\nu_j}-f_\nu^{eq})/f_\nu^{eq} $. In the last case 
the contributions to the collision integrals for
all the processes vanish for vanishing $\delta$, except for the
interactions of neutrinos with electrons, where the "driving force" term,
proportional to the temperature difference between $\nu$ and $e^\pm$, gives
a nonzero contribution. Moreover, for small $x<1$ this 
contribution is also very small because the temperature difference is 
approximately $T_\gamma/T_\nu \approx 1 + 0.005 x^{2.3}$ which 
is close to 1 for small x. Here we have defined $T_\nu = 1/a$.

However, in the case with the full distribution we erroneously omitted the 
factor
1.22 (from the Planck mass) for some of the reactions, which 
created  somewhat different results for the two
different approaches. Now, after we  corrected for this factor, 
the results are in perfect agreement.

In the recent paper~\cite{gnedin} the factor  2 in 
the rate of the reaction $\nu_a \nu_a \rightarrow \nu_a \nu_a $ is missing.
Fortunately, the difference in neutrino energy density due to this factor 
is very small, $\sim 10^{-6}$, which is well below the numerical precision. 

We can define the effective number of neutrino species at asymptotically
large time as:
\begin{equation}
N_{\mbox{eff}} =  \frac{\rho_{\nu_e}+2 \rho_{\nu_\mu}}{\rho_{\nu}^{eq}} 
\frac{\rho_\gamma^{eq}}{\rho_\gamma} ~,
\label{N_eff}
\end{equation}
where the photon energy density is
$\rho_\gamma=(\pi^2/15) (a T_{\gamma})^4$ and the 
equilibrium quantities are 
$\rho_{\nu}^{eq}=(7/8)(\pi^2/15)/a^4$ and $\rho_\gamma^{eq}=
(\pi^2/15)(a T_{\gamma}^{eq})^4$.

In the last two columns in tables 1 and 2 we present $N_{\mbox{eff}}$ 
from eq.~(\ref{N_eff}).
The error on  $N_{\mbox{eff}}$, $\delta N_{\mbox{eff}}$, 
comes from the numerical error in the definition
of $aT_\gamma^{eq}=1.40102$ (see section~\ref{timeevolv}).
We denote this error $\delta(aT_\gamma^{eq})$  in tables 1 and 2.

\begin{table}
\begin{center}
\begin{tabular}{|c||c|c|c|c|c|c|c|}
\hline
 
&&&&&\\
Program&points&
~~$a T_\gamma$~~&~~$\delta(a T_\gamma^{eq})$~~&~~$\delta\rho_{\nu_e}/\rho_{\nu_e}
 $~~&~~$\delta\rho_{\nu_{\mu}}/\rho_{\nu_{\mu}} 
$~~&$N_{\mbox{eff}}$&$\delta N_{\mbox{eff}}$\\
&&&&&&&\\
\hline
\hline
             &100&1.399130&0.000031&0.9435\%&0.3948\%&3.03392&-0.0003\\
$\delta(x,y)$&200&1.399135&0.000031&0.9458\%&0.3971\%&3.03395&-0.0003\\
             &400&1.399135&0.000031&0.9459\%&0.3972\%&3.03396&-0.0003\\
\hline
             &100&1.399079&-0.000024&0.9452\%&0.3978\%&3.03398&0.0003\\
$f(x,y)$     &200&1.399077&-0.000023&0.9459\%&0.3986\%&3.03401&0.0003\\
             &400&1.399077&-0.000023&0.9461\%&0.3990\%&3.03402&0.0003\\
\hline
\hline
\end{tabular}
\end{center}
\caption{Two ways of calculation.}
\end{table}

\subsection{Initial conditions}

Now, let us discuss the choice of the initial time $x_{in}$. We made 
the runs for the system of kinetic equations  with
three different values $x_{in}=0.1, 0.2$ and $0.5$. We found that the
results of the runs with $x_{in}=0.1$ and $x_{in}=0.2$ are the same with an
accuracy of $10^{-5}$. This 
means that for $x \le 0.2$ we can neglect the non-equilibrium corrections 
to the neutrino distribution functions.

Let us note that already at $x_{in}=0.1$ the dimensionless photon temperature 
differs from unity, $a \, T \neq 1$. For our calculations we took 
two possible sets of initial conditions. The first was used in 
the paper~\cite{dhs}: 
\begin{equation}
f_{\nu_{e (\tau)}} = f_{eq} = \frac{1}{e^{y} + 1} ~~,~~ a \,
T(x_{in}=0.1)=1.00006~. 
\label{in_1}
\end{equation}

These conditions correspond to separate energy conservation 
in the electromagnetic plasma
before the
time $x_{in}$. Note, that even though  $a T$  is very close to $1$, we 
need to keep $a T \neq 1$, because our precision for the 
equilibrium temperature
is of the order $0.00003$. The second set of initial conditions is similar 
to the one we used in the paper~\cite{massive}:

\begin{equation}
f_{\nu_{e (\tau)}} = \frac{1}{e^{y/T} + 1} ~~,~~ a \, T(x_{in}=0.1)=1.00003~. 
\label{in_2}
\end{equation}
These conditions correspond to the neutrinos being in thermal equilibrium
with the electromagnetic plasma before $x_{in}$. 
We found that both initial conditions give the same results for  the
neutrino energy density and other essential quantities for
$x_{in}=0.1$, i.e. the difference in the results is 
less than our numerical errors. 
If one instead would choose $x_{in}=0.2$ then the condition (\ref{in_2}) 
is more precise.

\subsection{Momentum grid}

We took the dimensionless momentum interval $0 \le y \le 20$.
For equilibrium neutrinos 
$\rho_\nu(y>20)/\rho_\nu \approx 3~\times~ 10^{-6}$.
Because the nonequilibrium correction to the 
neutrino energy density is of the order  
$1\%$, neglecting $y>20$ does not affect the result for the 
neutrino energy density
even if the calculation is done with $1\%$ precision of the effect.
There is a somewhat bigger correction in the reaction rates due to 
the preexponential factor, $ \sim p^2$, but even for that the
momentum cut-off at $y=20$ provides a sufficient accuracy.

In order to choose the distribution of the momentum grid properly 
let us take a look on the differential energy density of the neutrinos
$d\rho_\nu/dy = (1/\pi^2) y^3 f(y)$.
$97.5 \%$ of the energy density comes from  particles with momentum 
in the interval $1<y<10$. Particles with momentum $0.1<y<1$ give
$1.4 \%$ of the total energy density and particles with $10<y<20$
give $1.1 \%$ of the total energy density.
Also note that non-equilibrium corrections are particularly important
for particles with large momenta. 

These arguments  advocate  the use of a  linear distribution in the region
$0\le y \le 20$ or 
log(y) distribution in the region $0.1 \le y \le 20$.
We found that the difference between these  approaches with the same number 
of points in grid is about $10^{-6}$ for the neutrino energy density.
 
The authors of the paper~\cite{gnedin} chose log(y) distribution of  points 
in grid $10^{-5.5}<y<10^{1.7}$ with 40 points per decade. 
With such a choice  more than half of the points lie in the region
$y<0.1$, which gives only $0.0002 \%$ contribution to the 
neutrino energy density. 
In the most important decade $1<y<10$ they have only 40 points.

In order to check the errors connected with a finite number of points in 
grid we took the 100, 200 and 400 point grids. 
The results hereof are presented in table 1.

\subsection{ Time evolution}
\label{timeevolv}
We used three different methods of time evolution.

1) Euler method.  
We control the errors 
connected with a finite number of points
in time $x$ in the following way. First, we run the program
with some fixed number of points in $x$, distributed in the time interval
$x_{in}<x<x_f$ in such a way that the distribution functions do not change
significantly at any 
momentum point $y$ during one time iteration $dx$. Then we
run the 
program for the entropy conservation law (i.e. with equilibrium neutrinos)
with the same values of time $x_i$ as in the first run. Finally we compare
the asymptotical values of the temperature ratios with the 
theoretical value which is   
$T_\gamma/T_\nu=(11/4)^{1/3} = 1.40102$.
In order to  have good precision we require that the 
numerical error in these temperature ratio  should not be larger than 
$ \sim 0.00003$ (fourth column in table 1).

2) Bulirsch-Stoer method.
Instead of the simple time evolution we used the Bulirsch-Stoer method, 
described in the book~\cite{numrec}.

3) Method for stiff equations. 
In order to compare our results with the results  of the paper~\cite{gnedin}
we made calculations with their method for stiff equations. 

We found that the most precise (in calculation of $T_\gamma^{eq}$) 
is the Bulirsch-Stoer method,
but in the region of small time $x_{in}<x<1$ it requires 10 times more 
processor time than the Euler or the stiff method. Moreover, in the case 
that we take $x_{in} \ll 0.1$ only the stiff method takes a reasonably small 
number of time steps.

In the region of large time $x>1$ the situation is the  opposite.
The Bulirsch-Stoer method requires 10 times fewer time steps than  the Euler 
method. Unfortunately, we cannot control  the precision of  the 
stiff equations 
method in this time region. The problem is that the 
energy conservation law, which 
we use for  the 
evaluation of  the photon temperature, is {\em not} a stiff equation.    
In our calculations
with the stiff method we therefore evolved the photon
temperature as $T_\gamma = T_\gamma^{eq} + \delta T$.
The value of the equilibrium photon temperature $T_\gamma^{eq}$
is taken from the entropy conservation equation, while for the evolution 
of the small $\delta T$ we use the same time steps as we have for the kinetic 
equations in the stiff method.

\begin{table}
\begin{center}
\begin{tabular}{|c||c|c|c|c|c|c|c|}
\hline
 
&&&&&\\
Program&points&
~~$a T_\gamma$~~&~~$\delta(a T_\gamma^{eq})$~~&~~$\delta\rho_{\nu_e}/\rho_{\nu_e}
 $~~&~~$\delta\rho_{\nu_{\mu}}/\rho_{\nu_{\mu}} 
$~~&$N_{\mbox{eff}}$&$\delta N_{\mbox{eff}}$\\
&&&&&&&\\
\hline
\hline
             &&&&&&&\\
$\delta(x,y)$&100&1.399130&0.000031&0.9435\%&0.3948\%&3.03392&-0.0003\\
 Euler       &&&&&&&\\
\hline
             &&&&&&&\\
$f(x,y)$     &100&1.399079&-0.000024&0.9452\%&0.3978\%&3.03398&0.0003\\
Euler+BS     &&&&&&&\\
\hline
             &&&&&&&\\
$f(x,y)$     &100&1.399100&$10^{-7}$&0.9463\%&0.3981\%&3.03401&$10^{-6}$\\
 BS          &&&&&&&\\
\hline
             &&&&&&&\\
$f(x,y)$     &100&1.399060&&0.9518\%&0.3976\%&3.03440&\\
stiff        &&&&&&&\\
\hline
             &&&&&&&\\
$f(x,y)$     &100&1.399085&&0.9399\%&0.3934\%&3.03401&\\
stiff+BS     &&&&&&&\\
\hline
\hline
\end{tabular}
\end{center}
\caption{Different time evolution algorithms.
}
\end{table}

In the calculations with  the corrections, $\delta(x,y)$, to  the 
distribution functions 
we used  the Euler method.
In calculations with  the total distribution functions, $f(x,y)$,  we
used  the Euler method for $x<1$ and  the Bulirsch-Stoer method for $x>1$.
In order to compare our results with the paper~\cite{gnedin}, 
we also made  the calculations using the  stiff method in two different ways:
with the stiff method evolution for all $x$, and with a combination of the 
stiff method for $x<1$ and  the Bulirsch-Stoer method for $x>1$. 
In table 2 we compare the results for these ways of calculation
with 100 points grid. We found that all ways of calculation give 
an effective number of neutrino species around $N_{eff}=3.0340$,
except the stiff method which gives a slightly larger value 
$N_{eff}=3.0344$.

\section{Conclusion}
We have seen, that the effect of non-equilibrium neutrinos can be
calculated with a very good accuracy if one takes a large enough number
of momentum points 
in the important region $1 < y < 10$ and a precise 
algorithm for the time evolution. Finally
let us present our results for late times. 
Photon temperature: 
$(a T_\gamma)_{final} = 1.39910 \pm 0.00003~ $,
correction to the energy density of electron neutrino: 
$\delta\rho_{\nu_e}/\rho_{\nu_e}
  = 0.946 \pm  0.001~$,
correction to the energy density of muon neutrino: 
$\delta\rho_{\nu_\mu}/\rho_{\nu_\mu}
  = 0.398 \pm  0.001~$,
and effective number of neutrino species: 
$N_{eff} = 3.0340 \pm  0.0003~$. 

\bigskip

{\bf Acknowledgment.}
We thank Sergio Pastor for pointing out the error
of 1.22 in the early version of our program.

The work of AD and SH was supported in part by the Danish National Science 
Research Council through grant 11-9640-1 and in part by Danmarks 
Grundforskningsfond through its support of the Theoretical Astrophysical 
Center. 
The work of DS was supported in part by the Russian Foundation for 
Fundamental Research through grants 97-02-17064A and 98-02-17493A.

\end{document}